\documentclass[prb,twocolumn,showpacs,preprintnumbers,amsmath,amssymb]{revtex4}

\usepackage{graphicx}
\usepackage{dcolumn}
\usepackage{bm}

\newcommand{\Na}[1]{{\Hat{#1}}}
\newcommand{\Fe}[1]{{\bf #1}_{F}}
\newcommand{\Fep}[1]{{\bf #1}'_{F}}
\newcommand{\Avp}[1]{\left\langle{#1}\right\rangle_{\Fe p}}
\newcommand{\Avpp}[1]{\left\langle{#1}\right\rangle_{\Fep p}}

\begin{document}


\title{Disordered Josephson Junctions of $d$-Wave Superconductors}

\author{Thomas L\"uck}
\email{lueckt@physik.uni-augsburg.de}
\author{Peter Schwab}%
\author{Ulrich Eckern}
\affiliation{Institut f\"ur Physik, Universit\"at Augsburg,
D-86135 Augsburg, Germany}

\author{A. Shelankov}
\affiliation{Department of Theoretical Physics,
Ume{\aa} University, 901 87, Sweden
}%
\altaffiliation[]{A.~F. Ioffe Physico-Technical Institiute, 19021 St. Petersburg, Russia}

\date{\today}

\begin{abstract}
We study the Josephson effect between weakly coupled $d$-wave
superconductors within the quasiclassical theory, in particular, the
influence of interface roughness on the current-phase relation and the
critical current of mirror junctions and $45^\circ$ asymmetric
junctions.  For mirror junctions the temperature dependence of the
critical current is non-monotonic in the limit of low roughness, but
monotonic for very rough interfaces.  For $45^\circ$ asymmetric
junctions with a linear dimension much larger than the superconducting
coherence length we find a $\sin(2\varphi)$-like current-phase
relation, whereas for contacts on the scale of the coherence length or
smaller the usual $\sin\varphi$-like behavior is observed. Our results
compare well with recent experimental observations.

\end{abstract}

\maketitle

\section{Introduction}
It is commonly believed that high-$T_c$ superconductors are described
by an order parameter which dominantly exhibits $d$-wave
symmetry. This was especially established by SQUID-like experiments,
e.g. the corner-SQUID experiments by Wollmann et al.~\cite{Wo93} or
the tri-crystal experiment by Tsuei et al.~\cite{Ts94}, which provide
a phase-sensitive test of the order parameter. This observation was
followed by numerous theoretical studies of the Josephson effect
between $d$-wave superconductors with rather unexpected results; a
review can be found in Refs.~\cite{TaKa97,KaTa00,Lo01}.

For instance a mirror junction, where the order parameter on the left
and right hand side is rotated by the same angle but in opposite
directions, should exhibit a non-monotonic temperature dependence of
the critical current~\cite{Ba96,TaKa96,TaKa97}; this behavior is
related to a transition to a $\pi$-junction state at low
temperatures. For the $45^\circ$ asymmetric junction, in which the
order parameter is rotated only on one side by $45^\circ$, a
$\sin(2\varphi)$-like current-phase relation was
predicted~\cite{TaKa97}, as the leading contribution to the tunnel
current vanishes due to symmetry.  Experimentally, the temperature
dependence of the critical current~\cite{Di90,Hi98,Ar00} and the
current-phase relation~\cite{Il98,Il99,Il01} of artificially produced
grain boundaries with a well defined lattice orientation were
determined.  For mirror junctions the predicted non-monotonic behavior
of the temperature-dependent critical current was found by Il'ichev et
al.~\cite{Il01}, whereas in other cases a monotonic behavior as in
usual $s$-wave superconductors was reported~\cite{Di90,Hi98,Ar00}. In
the experiment on an asymmetric junction~\cite{Il99} not all samples
showed the predicted $\sin(2\varphi)$-like current-phase relation.

As the $d$-wave order parameter is sensitive to disorder scattering it
is believed that the origin for these differing results is interface
roughness: One possibility is the existence of facets with spatially
varying orientation which occur at the interfaces on a $\mu{\rm
m}$-scale~\cite{Ma93,Ma96}.  This kind of disorder can theoretically
be described by a model consisting of randomly oriented
mirrors~\cite{Th92,Fo97}. On the other hand roughness on the atomic
scale is present as well. In theoretical studies this is a non-trivial
problem which we will tackle in this work.  In previous studies
interface  roughness was modeled by a thin dirty layer next to a specular
interface~\cite{Ba96,Ba97,Po99}. We use an alternative approach
which allows us to study, at least in principle, also individual
realizations of the disorder; this might be of importance for
mesoscopicly small junctions.

For the calculations we will use the quasiclassical theory of
superconductivity~\cite{Ei68,RaSm86} which is valid for spatial
variations on scales which are large compared to the Fermi wave
length, $\lambda_F$; this is marginally fulfilled in the ${\rm
CuO_2}$-planes of high-$T_c$ materials as $\xi_0/\lambda_F\approx 5$
($\xi_0$: superconducting coherence length at zero temperature in the
${\rm CuO_2}$-planes). Near interfaces the quasiclassical theory must
be supplemented by boundary conditions. A specular interface can be
described by Zaitsev's~\cite{Za84} boundary conditions. We use a more
general scheme as suggested by Shelankov and Ozana~\cite{ShOz00}; this
approach is extended in order to find suitable boundary conditions for
interfaces. The properties of the interface, e.g. roughness or
transparency, are incorporated by a scattering matrix. To describe an
interface without regular structure we phenomenologically choose
random scattering matrices. A similar approach has already been
successfully applied to examine tunnel junctions between a normal
metal and a $d$-wave superconductor~\cite{Ya96,Lu01}.

The paper is arranged as follows: In the next section we briefly
introduce the quasiclassical theory. We then describe in detail the
boundary conditions used in this work, and discuss our choice of the
scattering matrix and its properties.  In section~\ref{RD} we first
present our results for mirror junction: We study the temperature
dependent critical current for interfaces with varying roughness and
make a quantitative comparison with experimental data.  We furthermore
examine the transition to a $\pi$-junction, where a spontaneous
current parallel to the junction is present. Secondly we present the
results for $45^\circ$ asymmetric junctions, where additionally to the
average quantities also statistical fluctuation of the critical
current are examined. In both cases we discuss in detail the influence
of interface roughness, which might lead to clear modifications of
the results for the clean case.  Concluding remarks are given in
section~\ref{CO}.

\section{Method}
In this section we introduce the main ingredients for our
phenomenological model of rough superconducting contacts: The
superconducting state is described by the quasiclassical theory, which
is supplemented by boundary conditions to take into account interface
scattering. We first recall the basic formalism.

\subsection{Theory of quasiclassical Green's functions}
The quasiclassical matrix Green's function is determined by the Eilenberger
equation which in thermal equilibrium reads
\begin{equation}\begin{split}\label{Eilen_eq}
\left[\Na\tau_3E+e\Na\tau_3\Fe v\cdot{\bf A}({\bf r})+
i\Na\Delta(\Fe p,{\bf r}),\Na g(E,\Fe p;{\bf r})\right]+\\
+i\Fe v\cdot\partial_{\bf r}\Na g(E,\Fe p;{\bf r})=0\;.
\end{split}\end{equation}
Here $\Na\tau_i$ are the Pauli matrices in Nambu space.  The
physically relevant solution is fixed by the normalization condition
\begin{equation}\label{norm}
\left[\Na g(E,\Fe p;{\bf r})\right]^2=\Na 1\;.
\end{equation}
The (spin-singlet) order parameter has the matrix form
\begin{equation}
\hat \Delta(\Fe p,{\bf r})=
\begin{pmatrix}
0 & \Delta(\Fe p,{\bf r})\\ \Delta^*(\Fe p,{\bf r})& 0
\end{pmatrix}
\end{equation}
and must be determined self-consistently via
\begin{equation}\label{OP}
\Na\Delta(\Fe p,{\bf r})=-\pi{\cal N}_0  T\sum_{|E_n|<E_c}
\Avpp{V(\Fe p,\Fep p)\Na g(iE_n,\Fep p;{\bf r})},
\end{equation}
where $E_n=\pi T(2n+1)$ are the Matsubara energies ($\hbar=k_{B}=1$)
and ${\cal N}_0$ is the density of states (DOS) at the Fermi level in
the normal state. For convenience we assume a cylindrical Fermi
surface; $\Avp{\dots}$ denotes the Fermi surface average. In the BCS
approximation the pairing interaction is given by its strength $V$,
its cut-off energy $E_c$, and its direction dependence which can be
expanded for a $d$-wave order parameter as follows
\begin{equation}
V(\Fe p,\Fep
p)=V\eta(\Fe p)\eta(\Fep p)
\end{equation}
with
\begin{equation}
\eta(\Fe p)=(p_{F,x}^2-p_{F,y}^2)/p_F^2\;.
\end{equation}
For a homogeneous system and temperature $T=0$ one finds
$\Na\Delta(\Fe p,{\bf r})=\Na\tau_1\Delta_0\eta(\Fe p)$, with
$\Delta_0\approx 2.14T_c$.

Having solved these equations for the Green's function, various
observables can be calculated. In this work the current density (in
thermal equilibrium) is of particular interest:
\begin{equation}\label{current}
{\bf j}({\bf r})=-ie\pi{\cal N}_0 T\sum_{n=-\infty}^{\infty}
{\rm Tr}\left[\Na\tau_3\Avp{\Fe v \Na g(iE_n,\Fe p;{\bf r})}\right]\;.
\end{equation}
Another important quantity is the DOS which is given by
\begin{equation}
{\cal N}(E,{\bf r})=\Avp{{\cal N}(E,\Fe p,{\bf r})}\;,
\end{equation}
with the angle-resolved DOS
\begin{equation}
{\cal N}(E,\Fe p,{\bf r})= \frac{1}{2}{\cal N}_0{\rm Re}\big\{{\rm
Tr}\left[\Na\tau_3\Na g(E+i0_+,\Fe p;{\bf r})\right]\big\}\;.
\end{equation}
In order to solve the Eilenberger equation it is useful to
parameterize the Green's function as suggested in
Refs.~\cite{Na93,ScMa95,ShOz00}
\begin{equation}\label{para}
\Na g=\frac{1}{1-ab}
\begin{pmatrix}
1+ab & -2a\\
2b & -(1+ab)
\end{pmatrix}.
\end{equation}
Note that the normalization condition is fulfilled by construction.
The ansatz~(\ref{para}) yields the Riccati type equations for the
amplitudes $a(E,\Fe p;{\bf r})$ and $b(E,\Fe p;{\bf r})$:
\begin{align}
\label{ricca}
\Fe v\cdot\partial_{\bf r}a&=\Delta^* a^2+2i(E+e\Fe v\cdot{\bf A})a-
\Delta\;,\\
\label{riccb}
\Fe v\cdot\partial_{\bf r}b&=\Delta b^2-2i(E+e\Fe v\cdot{\bf A})b-
\Delta^*\;.
\end{align}
The (numerical) evaluation of these amplitudes along classical
trajectories ${\bf r}(\lambda)={\bf r}_0+\lambda\Fe v/v_F$ is more
convenient than solving the Eilenberger Eq.~(\ref{Eilen_eq}) together
with the normalization condition Eq.~(\ref{norm}). In the case ${\rm
Im}[E]>0$, i.e. for the retarded Green's function or the Matsubara
Green's function with $E_n>0$, the integration of Eq.~(\ref{ricca}) is
stable in positive $\Fe v$-direction, and Eq.~(\ref{riccb}) is stable
in negative $\Fe v$-direction; the directions must be reversed for
${\rm Im}[E]<0$. In the cases considered here the trajectories start and
end in the bulk ($\lambda=\pm\infty$) where the order parameter has no
spatial dependence; then Eqs.~(\ref{ricca}) and~(\ref{riccb}) yield the
homogeneous solutions ($\partial_{\bf r}a=0$, $\partial_{\bf r}b=0$)
\begin{align}
\label{a_ini}
&a(E,\Fe p;\lambda\to-\infty)=\frac{-i\Delta_-(\Fe p)}
{E+\sqrt{E^2-|\Delta_-(\Fe p)|^2}},\\
\label{b_ini}
&b(E,\Fe p;\lambda\to+\infty)=\frac{i\Delta_+^*(\Fe p)}
{E+\sqrt{E^2-|\Delta_+(\Fe p)|^2}},
\end{align}
where $\Delta_\pm(\Fe p)$ is the order parameter for
$\lambda\to\pm\infty$. Starting from these initial values
Eqs.~(\ref{ricca}) and~(\ref{riccb}) must be integrated in positive
and negative directions, respectively. It is then possible to construct
the Green's function on the trajectory via Eq.~(\ref{para}).

In the following subsection we show that this parameterization is a good
starting point for implementing boundary conditions which are
necessary to study interfaces.

\subsection{Boundary conditions}
\begin{figure}[b]
\includegraphics[width=0.35\linewidth]{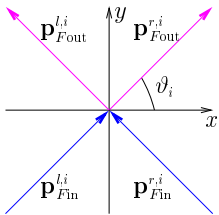}
\caption{\label{gridil}At an specular interface the Green's functions on
the four trajectories with the same parallel momentum $p_{F,y}$ are
coupled coherently.}
\end{figure}

The amplitudes $a$ with $\Fe p$ pointing towards the interface
(in-direction) can be calculated by integrating Eq.~(\ref{ricca}) with
the initial value in the bulk given by Eq.~(\ref{a_ini}); analogously
the $b$'s with $\Fe p$ pointing away from the interface
(out-direction) can be calculated via Eqs.~(\ref{riccb})
and~(\ref{b_ini}).  The boundary conditions provide the $a$'s in
out-direction and the $b$'s in in-direction directly at the interface;
starting with these values the integration of Eqs.~(\ref{ricca})
and~(\ref{riccb}), respectively, yields the $a$'s in the out- and the
$b$'s in the in-direction.

For simplicity we consider only a finite number $n$ of trajectories on
each side of the junction, i.e. ${\bf p}^{s}_{F\rm in/out}\to{\bf
p}^{s,i}_{F{\rm in/out}}$ ($i=1,\dots,n$, $s=l/r$: left/right); we
choose the numbering of the directions such that (see
Fig.~\ref{gridil})
\begin{equation}\label{parmom}
{\bf p}^{l,i}_{F{\rm in},y}={\bf p}^{l,i}_{F{\rm out},y}=
{\bf p}^{r,i}_{F{\rm in},y}={\bf p}^{r,i}_{F{\rm out},y}\;.
\end{equation}
We denote the amplitude $a$ on the $i^{\rm th}$ in/out-trajectory on
the side $s$ of the junction by $a^{s,i}_{\rm in/out}$, and
$b^{s,i}_{\rm in/out}$ analogously. To formulate the boundary
conditions we follow Ref.~\cite{ShOz01} by introducing the determinant
${\cal D}$ as follows:
\begin{equation}
{\cal D}\left(\{a^{s,i}_{\rm in}\},\{b^{s,i}_{\rm out}\}\right)=
\det\left({\bf 1}-{\bf S}{\bf a}{\bf S}^\dagger{\bf b}\right)\;;
\end{equation}
the $2n\times 2n$ diagonal matrices ${\bf a}$ and ${\bf b}$ read
\begin{align}
&{\bf a}={\rm diag}\left[\{a^{l,i}_{\rm in}\},\{a^{r,i}_{\rm in}\}\right],\\
&{\bf b}={\rm diag}\left[\{b^{l,i}_{\rm out}\},\{b^{r,i}_{\rm out}\}\right],
\end{align}
where the $a$'s and $b$'s are the amplitudes at the interface
($x=0$). The unitary $2n\times 2n$ scattering matrix ${\bf S}$, which
contains the physical properties of the interface, has the block
structure
\begin{equation}\label{S-matrix}
{\bf S}=\begin{pmatrix}
{\bf S}^{ll}&{\bf S}^{lr}\\{\bf S}^{rl}&{\bf S}^{rr}
\end{pmatrix},
\end{equation}
where the elements are $n\times n$ matrices.
As described in detail in~\cite{ShOz01} we define the quantities
\begin{align}
&{\cal A}^{s,i}_{0}={\cal D}\Big|_{a^{s,i}_{{\rm in}}=0}\;,
&{\cal A}^{s,i}_{1}=
\frac{\partial{\cal D}}{\partial a^{s,i}_{{\rm in}}}
\Big|_{a^{s,i}_{{\rm in}}=0}\;,\\
&{\cal B}^{s,i}_{0}={\cal D}\Big|_{b^{s,i}_{{\rm out}}=0}\;,
&{\cal B}^{s,i}_{1}=
\frac{\partial{\cal D}}{\partial b^{s,i}_{{\rm out}}}
\Big|_{b^{s,i}_{{\rm out}}=0}\;.
\end{align}
The boundary conditions yield the unknown values of $a$ and
$b$ at the interface via
\begin{equation}
a^{s,i}_{{\rm out}}=-\frac{{\cal B}^{s,i}_{1}}{{\cal B}^{s,i}_{0}},
\qquad
b^{s,i}_{{\rm in}}=-\frac{{\cal A}^{s,i}_{1}}{{\cal A}^{s,i}_{0}}.
\end{equation}
To apply these boundary conditions to interfaces, we must ensure
current conservation across the junction.  As discussed in
Ref.~\cite{ShOz01} the above boundary conditions yield the following
conservation law:
\begin{equation}\label{C_con}\begin{split}
\sum_{i=1}^n\big[
&\tilde j^l(E,{\bf p}^{l,i}_{F\rm in})-\tilde j^l(E,{\bf p}^{l,i}_{F\rm out})
\big]\\
&=\sum_{i=1}^n\big[
\tilde j^r(E,{\bf p}^{r,i}_{F\rm out})-\tilde j^r(E,{\bf p}^{r,i}_{F\rm in})
\big]
\end{split}\end{equation}
where the contribution of each direction is given by
\begin{equation}
\tilde j^s(E,{\bf p}_{F})={\rm Tr}
\left[\Na\tau_3\Na g^s(E,{\bf p}_{F};0)\right].
\end{equation}
On the other hand, we impose current conservation perpendicular to the
junction (see Eq.~(\ref{current}))
\begin{equation}\label{C_con_con}\begin{split}
\big\langle v_{F,x}&\left[
\tilde j^l(E,{\bf p}^l_{F\rm in})-\tilde j^l(E,{\bf p}^l_{F\rm out})
\right]\big\rangle_{{\bf p}^l_{F\rm in}}\\
&=\big\langle v_{F,x}\left[
\tilde j^r(E,{\bf p}^r_{F\rm in})-\tilde j^r(E,{\bf p}^r_{F\rm out})
\right]\big\rangle_{{\bf p}^r_{F\rm in}}\;.
\end{split}\end{equation}
Note that the directions ${\bf p}^{s,i}_{F\rm in}$ and ${\bf
p}^{s,i}_{F\rm out}$ are related by Eq.~(\ref{parmom}), see
Fig.~\ref{gridil}.  To ensure current conservation perpendicular to
the junction, we construct the grid of the discrete directions so that
the term $v_{Fx}=v_Fp_{F,x}/p_F$ is already taken into account. This
is guaranteed by choosing the grid as
\begin{equation}\label{grid}\begin{split}
&{\bf p}^{l,i}_{F\rm in}=p_F
\begin{pmatrix}\cos\vartheta_i\\\sin\vartheta_i\end{pmatrix},\qquad
{\bf p}^{r,i}_{F\rm in}=p_F
\begin{pmatrix}-\cos\vartheta_i\\\sin\vartheta_i\end{pmatrix},\\
&{\bf p}^{l,i}_{F\rm out}=p_F
\begin{pmatrix}-\cos\vartheta_i\\\sin\vartheta_i\end{pmatrix},\qquad
{\bf p}^{r,i}_{F\rm out}=p_F
\begin{pmatrix}\cos\vartheta_i\\\sin\vartheta_i\end{pmatrix},\\
&\sin\vartheta_i=\frac{2i}{n+1}-1,\qquad i=1,\dots,n.
\end{split}\end{equation}
In other words this grid takes into account that the rate of
scattering events at the interface decreases with smaller angles
of incidence.

The probability of scattering from the direction ${\bf p}^{s,j}_{F\rm
in}$ into an interval $[{\bf p}^{s',i}_{F\rm out},{\bf
p}^{s',i+1}_{F\rm out}]$ ($s,s'=l/r$) reads
\begin{equation}
P_{s's}(\vartheta_j\to\vartheta_i)\Delta\vartheta_i=|S_{ij}^{s's}|^2\;'
\end{equation}
with $\Delta\vartheta_i=\vartheta_i-\vartheta_{i-1}$. Using the
non-equidistant grid as defined in Eq.~(\ref{grid}) for a large
number of scattering channels, $n\gg 1$, we find
\begin{equation}\label{SPD}
P_{s's}(\vartheta_j\to\vartheta_i)=
\frac{n}{2}\cos\vartheta_i|S_{ij}^{s's}|^2\;.
\end{equation}
In the next step one has to find the appropriate scattering matrices
for a given physical realization. In the following subsection we study the case
of a specular and an irregular rough interface.

\subsection{\label{SM}Scattering matrix}
The most simple case is a specular interface where only trajectories
with the same parallel momentum are coupled. Then the block matrices
that form the scattering matrix in Eq.~(\ref{S-matrix}) have diagonal
form:
\begin{equation}\begin{split}\label{S_ideal}
&{\bf S}^{ll}=-{\bf S}^{rr}={\bf R}=
{\rm diag}\left[\sqrt{1-{\cal T}(\vartheta_i)}\right],\\
&{\bf S}^{lr}={\bf S}^{lr}={\bf T}=
{\rm diag}\left[\sqrt{{\cal T}(\vartheta_i)}\right];
\end{split}\end{equation}
in particular, for a $\delta$-like boundary potential the direction-dependent
transparency reads~\cite{Br90}
\begin{equation}\label{trans}
{\cal T}(\vartheta)=
\frac{{\cal T}_0\cos^2\vartheta}{1-{\cal T}_0\sin^2\vartheta}\;,
\qquad{\cal T}_0\in[0,1].
\end{equation}
For a normal metal this leads to the resistance of the tunnel-barrier
\begin{equation}\label{res}
R_N^{-1}=\frac{4Ae^2{\cal N}_0v_F}{3\pi}{\cal T}_0\;,
\end{equation}
where $A$ is the area of the contact~\cite{Lu01}.

In the superconducting case the boundary conditions for a specular
interface can be solved for each direction individually, with the
result~{\cite{ShOz00,Es00}}
\begin{align}
\label{Za_a}
&a^{s,i}_{\rm out}=\frac{a^{\bar s,i}_{\rm in}
{\cal T}_i(1-a^{s,i}_{\rm in}b^{\bar s,i}_{\rm out})+
a^{s,i}_{\rm in}
{\cal R}_i(1-a^{\bar s,i}_{\rm in}b^{\bar s,i}_{\rm out})}
{{\cal T}_i
(1-a^{s,i}_{\rm in}b^{\bar s,i}_{\rm out})+
{\cal R}_i(1-a^{\bar s,i}_{\rm in}b^{\bar s,i}_{\rm out})}\;,\\
\label{Za_b}
&b^{s,i}_{\rm in}=\frac{b^{\bar s,i}_{\rm out}
{\cal T}_i(1-a^{\bar s,i}_{\rm in}b^{s,i}_{\rm out})+
b^{s,i}_{\rm out}
{\cal R}_i(1-a^{\bar s,i}_{\rm in}b^{\bar s,i}_{\rm out})}
{{\cal T}_i(1-a^{\bar s,i}_{\rm in}b^{s,i}_{\rm out})+
{\cal R}_i(1-a^{\bar s,i}_{\rm in}b^{s,i}_{\rm out})}\;
\end{align}
where $s=l/r$, $\bar s=r/l$, respectively, and ${\cal T}_i={\cal
T}(\vartheta_i)$, ${\cal R}_i=1-{\cal T}_i$.

A rough surface, on the other hand, is described by a scattering
matrix that does not only depend on the transparency but also on a
roughness parameter. Our starting point to construct such a unitary
scattering matrix is~\cite{Be97}
\begin{equation}\label{S_carlo}
{\bf S}=\begin{pmatrix} {\bf U}_1 &  0 \\ 0 & {\bf U}_2\end{pmatrix}
\begin{pmatrix} {\bf T} & {\bf R} \\ {\bf R} & -{\bf T}\end{pmatrix}
\begin{pmatrix} {\bf U}_3 & 0 \\ 0 & {\bf U}_4\end{pmatrix},
\end{equation}
with ${\bf R}$ and ${\bf T}$ as defined in Eq.~(\ref{S_ideal}), which
contain the information on the transparency. The unitary matrices
${\bf U}_k$ describe the interface roughness; for ${\bf U}_k={\bf 1}$
the scattering matrix for a specular interface obviously is recovered.

\begin{figure}[t]
\includegraphics[width=\linewidth]{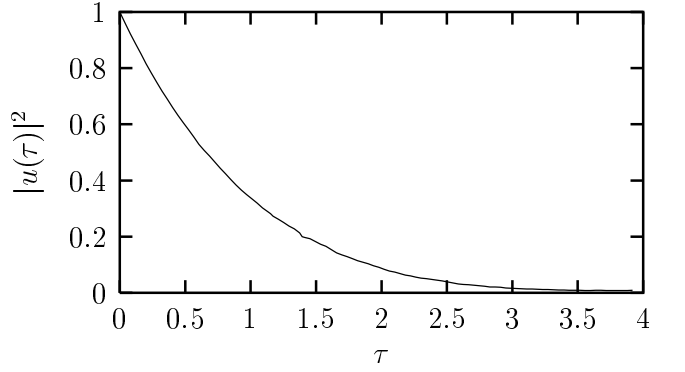}
\caption{\label{SSW}The weight for specular scattering, $|u|^2$,
decreases with growing $\tau$; for $\tau\gtrsim 3$ specular scattering
is almost completely suppressed ($|u|^2\ll 1$).}
\end{figure}

We will focus on interfaces without regular structure. To take into
account the statistical character of rough interfaces we choose the
${\bf U}_k$ to be unitary random matrices~\cite{Ya96,Lu01},
\begin{equation}
{\bf U}_k=\exp\{i{\bf H}_k\},
\end{equation}
where each ${\bf H}_k$ is a random matrix with Gaussian correlations:
\begin{equation}
\langle H_{k,ij}\rangle=0\;,\qquad
\langle {H_{k,ij}}^* H_{k',i'j'}\rangle=\frac{\tau}{2n}
\delta_{ii'}\delta_{jj'}\delta_{kk'}\;.
\end{equation}
Here $\langle\dots\rangle$ denotes the disorder average.  In the
following we use the abbreviations
\begin{equation}
|u(\tau)|=\langle|U_{k,ii}|^2\rangle\;,\qquad
|v(\tau)|=\langle|U_{k,i\neq j}|^2\rangle
\end{equation}
to describe the average
scattering probability; due to the unitarity it follows that
$|u|+(n-1)|v|=1$. Obviously for $\tau=0$ we find $|u(\tau)|=1$,
whereas for increasing (positive) $\tau$, we obtain a reduced value,
$|u(\tau)|<1$, as can be seen in Fig.~\ref{SSW}. A detailed discussion
can be found in~\cite{Lu01}.

Using Eq.~(\ref{SPD}) we find the average transmission probability
density for scattering from the direction ${\bf p}^{s,j}_{F\rm in}$ on
one side of the junction to ${\bf p}^{\bar s,i}_{F\rm out}$ on the other
side,
\begin{equation}\label{SPDi}\begin{split}
\langle P_{s\bar s}&(\vartheta_j\to\vartheta_i)\rangle=
\frac{\cos\vartheta_i}{2}
\Big\{n|u|^2{\cal T}(\vartheta_i)\delta_{ij}+\kappa(1-|u|)^2\\
&+|u|(1-|u|)[{\cal T}(\vartheta_i)+{\cal
T}(\vartheta_j)](1-\delta_{ij})\Big\}\;,
\end{split}\end{equation}
with
\begin{equation}
\kappa=\frac{1}{n}\sum_i{\cal T}(\vartheta_i)
\to\begin{cases}
{\cal T}_0/2 \qquad{\rm for}\quad{\cal T}_0\ll 1\\
1\qquad\qquad{\rm for}\quad{\cal T}_0=1\quad.
\end{cases}
\end{equation}
Furthermore the reflection probabilities $\langle P_{ss}\rangle$ can
easily be obtained by substituting ${\cal T}\to (1-{\cal T})$ and
$\kappa\to(1-\kappa)$ in Eq.~(\ref{SPDi}).  The continuum limit of
Eq.~(\ref{SPDi}) reads
\begin{equation}\begin{split}
&\langle P_{s\bar s}(\vartheta\to\vartheta')\rangle=
|u|^2{\cal T}(\vartheta')\delta(\vartheta'-\vartheta)+\\
&+\frac{\cos\vartheta'}{2}\Big\{\kappa(1-|u|)^2+
|u|(1-|u|)[{\cal T}(\vartheta')+{\cal T}(\vartheta)]\Big\}\;;
\end{split}\end{equation}
note that the weight of specular scattering is determined by the quantity
$|u|^2$: For a specular interface ($\tau=0$) it is $|u|^2=1$, whereas
in the very rough limit ($\tau\gg 1$) we find $|u|^2\to 0$ (see
Fig.~\ref{SSW}).
 
So far we discussed the boundary conditions at a single point of an
interface for a discrete number of directions $\Fe p$. A real
interface will have a finite cross section and the directions will be
continuous. Therefore the question arises how these boundary
conditions can be applied to real interfaces.

First we consider the discretization of the Fermi surface. A finite
number $n$ sets the typical angle, $\vartheta_c\approx\pi/n$, over
which the scattering probability can vary. For small $\vartheta_c$ the
fluctuations of the scattering probability with varying directions are
averaged out more effectively by the Fermi surface average that must
be taken to evaluate observables such as the current or the order
parameter. Therefore the self-averaging character of the physical
quantities increases with decreasing $\vartheta_c$; i.e. the
statistical fluctuations of physical quantities are diminished for a
small $\vartheta_c$.  It is important to note that only the
statistical fluctuations of physical quantities depend on the value of
$\vartheta_c$, but that the mean values are independent of
$\vartheta_c$, as long as $\vartheta_c\ll\pi$. Throughout the
following calculations we choose $n=40$.

Second we comment on the number of different realizations of the
scattering matrix that must be taken into account.  In the original
formulation of boundary conditions by Shelankov and
Ozana~\cite{ShOz00} the spatial resolution is limited by the Fermi
wave length, $1/p_F$; therefore the number of different realizations
should be of the order $Ap_F^2$, where $A$ is the area of the
contact. On the other hand, we expect that quantities like the
current density or the order parameter, which are of interest to us,
vary on the scale of the coherence length, $\xi_0$. Therefore we
choose one effective scattering matrix ${\bf S}$ to describe an area
$\xi_0^2$ of the whole contact. As the order parameter relaxes on the
coherence length it is reasonable to treat each of these areas
individually neglecting the influence from other part of the contact;
this approximation considerably simplifies the self-consistent
treatment of the quasiclassical theory. Finally the current
perpendicular to a junction of area $A$ is calculated via
\begin{equation}
I_x=\int\limits_A{\rm d}{A} j_x({\bf r})\approx
\xi_0^2\sum_{i=1}^{A/\xi_0^2}j_{x,i}\;,
\end{equation}
where each $j_{x,i}$ is evaluated for one particular realization of
the scattering matrix. For large junctions ($A\gg\xi_0^2$)
the current is given by the average over many scattering matrices; for
smaller junctions ($A$ of the order of $\xi_0^2$ or even smaller)
statistical fluctuations become relevant.

\section{\label{RD}Results and Discussion}
\begin{figure}[b]
\includegraphics[width=\linewidth]{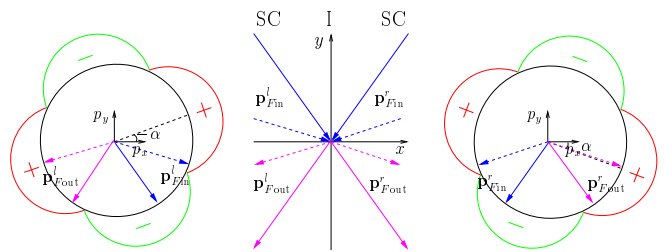}
\caption{\label{IDMJ}For mirror junctions the order parameter on the
left and the right hand side of the junction are rotated by the same
angle $\alpha$ but in opposite directions. For specular scattering the
order parameters on both in-trajectories and out-trajectories,
respectively, are equal.}
\end{figure}
\begin{figure}[b]
\includegraphics[width=\linewidth]{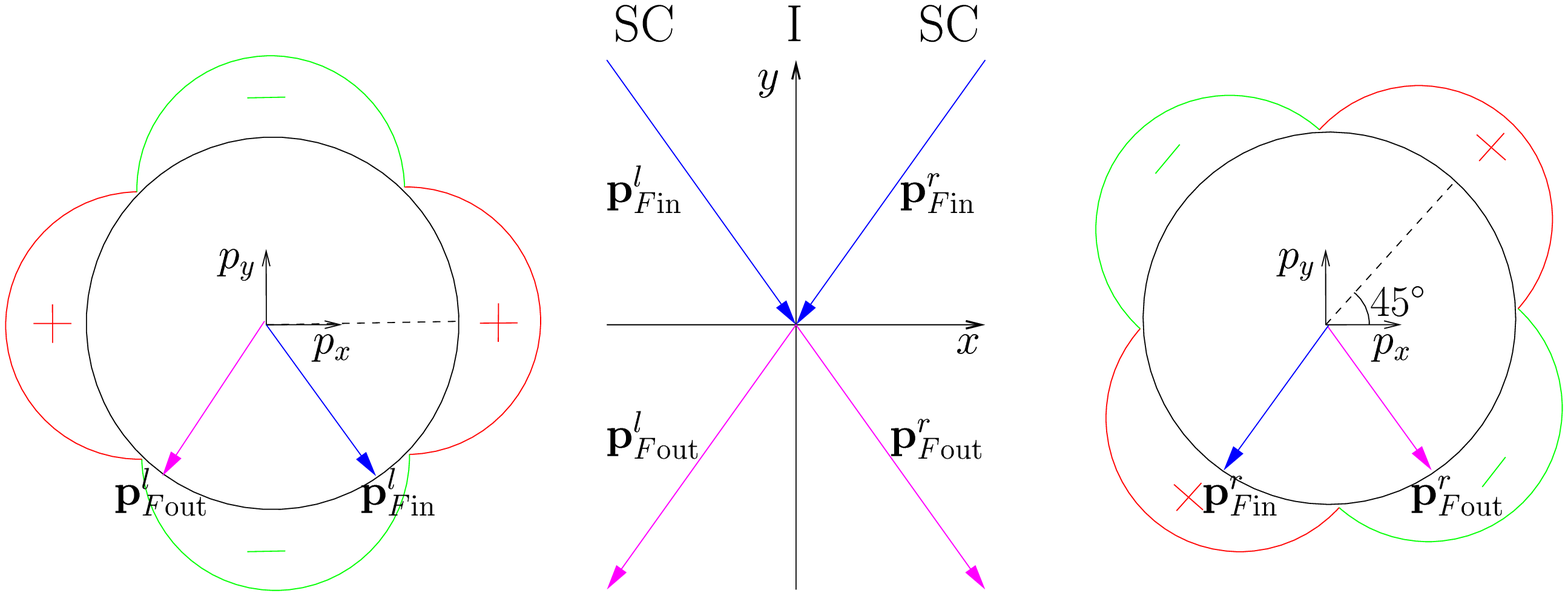}
\caption{\label{IDAJ}For a $45^\circ$ asymmetric junction the order
parameter is tilted on one side only.}
\end{figure}

We consider two kinds of Josephson junctions. First we discuss
so-called mirror junctions where the order parameter on both sides is
rotated by the same angle $\alpha$ but in opposite directions (see
Fig.~\ref{IDMJ}). Second we examine the $45^\circ$ asymmetric junction
where the order parameter is rotated only on the right hand side (see
Fig.~\ref{IDAJ}).

We recall the general expansion of the current-phase
relation between two superconductors,
\begin{equation}\label{I_exp}
I_x(T,\varphi)=I_1(T)\sin\varphi+I_2(T)\sin(2\varphi)+\dots
\end{equation}
with the superconducting phase difference, $\varphi$, between the left
and the right hand side.  It is also worth mentioning that the
current-phase can be obtained from the free energy of the junction via
\begin{equation}\label{freen}
I_x(T,\varphi)=2|e|\frac{\partial}{\partial\varphi}{\cal F}(T,\varphi).
\end{equation}
Besides their temperature dependence, the contributions $I_k$ scale as
\begin{equation}\label{I1_sca}
I_k\propto {\cal T}_0^k
\end{equation}
for small transparencies.  For $s$-wave superconductors and small
enough ${\cal T}_0$ only the linear term in the transparency, $I_1$,
is relevant.  In the following we will show that this needs not be
the case for $d$-wave superconductors.

\subsection{\label{MJ}Mirror junction}
\begin{figure}[t]
\includegraphics[width=\linewidth]{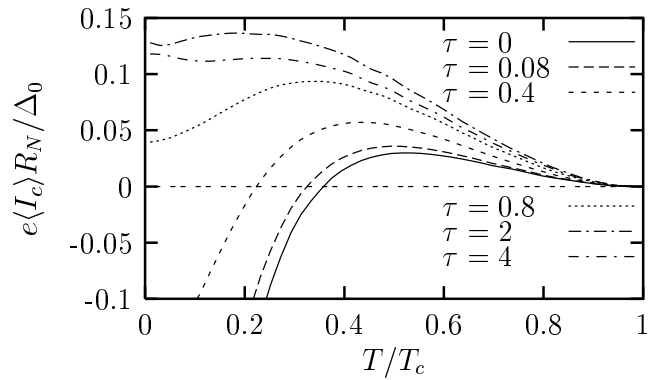}
\caption{\label{TDCC}Critical current for a mirror junction with
${\cal T}_0=0.01$ and $\alpha=22.5^\circ$. Negative values indicate a
$\pi$-junction state. Near $T=T_\pi$ the plotted curves make a jump
from a positive to a negative value and hence the critical current
stays finite. Due to the small transparency the jump is not visible in
this plot but in Fig.~\ref{CPR}.}
\end{figure}
\begin{figure}[t]
\includegraphics[width=\linewidth]{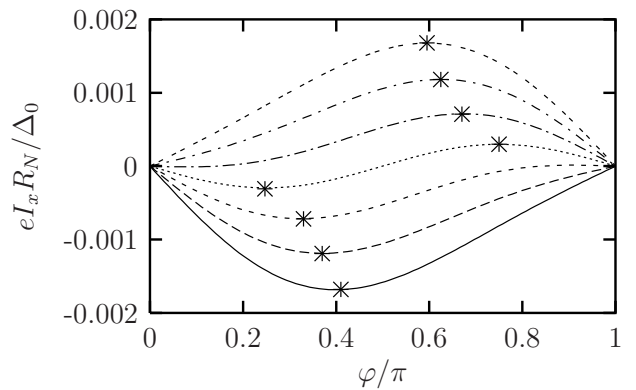}
\caption{\label{CPR}Current phase relation of a mirror junction
without roughness (${\cal T}_0=0.01$, $\alpha=22.5^\circ$) near the
temperature where it becomes a $\pi$-junction ($T=T_\pi\pm
k\cdot10^{-3}T_c$, $k=0,1,2,3$, increasing $T$ from bottom to top).
The asterisks mark the points which define the critical current. Note
that the critical current remains finite at $T=T_\pi$ due to the
$\sin(2\varphi)$-like contribution.}
\end{figure}
\begin{figure}[t]
\includegraphics[width=\linewidth]{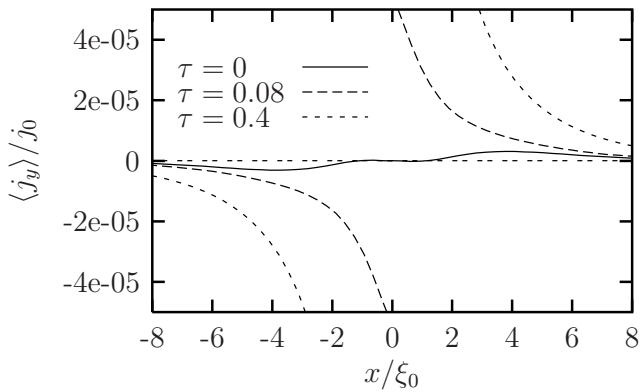}
\caption{\label{PC}Ground state current in $y$-direction at a
mirror junction with $\alpha=22.5^\circ$ and ${\cal T}_0=0.01$ at
$T=T_\pi$, i.e. $\varphi\approx\pi/2$. The current density is
increasing with the roughness of the interface.}
\end{figure}
\begin{figure}[t]
\includegraphics[width=\linewidth]{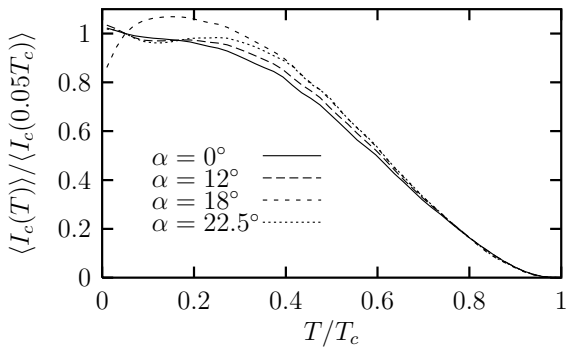}
\caption{\label{UNC}Critical current of mirror junction with different
orientation for very rough interfaces ($\tau=4$). The temperature
dependence of $I_c(T)/I_c(0.05T_c)$ is monotonic (within the numerical
error) and its angle dependence is only weak.}
\end{figure}

In this section we will consider mirror junctions, which are
schematically illustrated in Fig.~\ref{IDMJ}. For a junction with
purely specular scattering, a particular symmetry is present, which
reads
\begin{equation}\begin{split}
&\Delta^l({\bf p}^{l}_{F\rm in})=\Delta^r({\bf p}^{r}_{F\rm in})\;,\\
&\Delta^l({\bf p}^{l}_{F\rm out})=\Delta^r({\bf p}^{r}_{F\rm out})\;.
\end{split}\end{equation}
This means that for a given incoming trajectory the order parameter is
identical for the transmitted and the reflected quasi-particle.  All
directions can therefore be divided into two classes: (i) For some
directions the sign of the order parameter is the same for all
involved trajectories ($\Delta^l({\bf p}^{l}_{F\rm in})\Delta^l({\bf
p}^{l}_{F\rm out})>1$).  The behavior of these directions is similar
to that of usual $s$-wave superconductors: Their contribution to $I_1$
is positive and finite for $T<T_c$.  (ii) For the other directions the
sign for the in- compared to the out-trajectories changes
($\Delta^l({\bf p}^{l}_{F\rm in})\Delta^l({\bf p}^{l}_{F\rm
out})<0$). This has two crucial consequences: These directions have a
negative contribution to $I_1$ which moreover diverges in the tunnel
limit as $1/T$ for $T\to 0$. The reason for this divergence is the
existence of Andreev bound states near zero energy at the interface as
the reflected quasi-particles acquire a sign change of the order
parameter~\cite{Lo01,Lu01}; the divergence is cut-off by a finite
transparency, which leads to a shift of the bound state to finite
energies, and by disorder, which leads to a broadening of the level.

Altogether the $s$-wave-like contributions dominate for high
temperatures, which, for a fixed phase difference $0<\varphi<\pi$,
leads to a positive current, whereas for low temperature the anomalous
contributions are enhanced, which results in a negative current. In
other words, the ground state of the junction shifts from $\varphi=0$
($0$-junction) to $\varphi=\pi$ ($\pi$-junction) when decreasing the
temperature.  This transition occurs at the temperature $T_\pi$,
where both contributions cancel; at this temperature, the leading
$\sin\varphi$-like contribution vanishes ($I_1=0$) and the junction is
dominated by higher order terms.  As the amount of directions
preferring a $\pi$-junction increases with a growing angle $\alpha$,
also the temperature $T_\pi$ increases with $\alpha$.

In the following we will discuss the influence of interface roughness
on the temperature dependence of the critical current.  We calculate
the current contributions $I_1$ and $I_2$ as defined in
Eq.~(\ref{I_exp}); higher order contributions, $I_k$, $k>2$, can be
neglected as we consider only small transparencies. We determine the
critical current, $I_c$, from the absolute maximum of the current
phase relation. In the graphs a $\pi$-junction behavior is indicated
by a negative value. It should be mentioned that in experiments the
absolute value, $|I_c|$ is measured, thus a minimum in the temperature
dependent critical current is observed at $T=T_\pi$. We focus on the
case $\alpha=22.5^\circ$; for the presented results the transparency
is set to ${\cal T}_0=0.01$. The average current is evaluated by using
$20$ realizations of the scattering matrix; the statistical error is
less than $10\%$.

First, we concentrate on the clean case. There, as mentioned above,
the directions preferring a $\pi$-junction and those preferring a
$0$-junction compete: From the temperature dependence of the critical
current, as shown in Fig.~\ref{TDCC}, we find a $\pi$-junction
behavior below $T_\pi\approx 0.36\;T_c$, whereas for high temperatures
the $0$-junction state is favored.  Recalling Eq.~(\ref{freen}), the
shift of the ground state from $\varphi=0$ to $\varphi=\pi$ can be
seen in the current-phase relation for decreasing temperature near
$T_\pi$ (see Fig.~\ref{CPR}). In particular a finite critical current
can be observed even for $T=T_\pi$, since the contribution $I_2$
remains finite, whereas $I_1$ vanishes. In Fig.~\ref{TDCC} the jump in
the current at $T=T_\pi$ is too small to be visible; the reason is the
tiny transparency (${\cal T}_0=0.01$) we used in this calculation.
Moreover it can be seen in Fig.~\ref{CPR} that the current-phase
relation at $T_\pi$ is $\sin(2\varphi)$-like, which is related to the
degeneracy of the ground state. The ground state exhibits a
non-trivial phase difference $\varphi=\pi/2,3\pi/2$ which leads to a
current parallel to the interface; notice that also the anisotropy of
the order parameter plays a crucial role such that the contributions
with opposite $y$-component do not cancel each other. The parallel
current is presented in Fig.~\ref{PC}, where we use $j_0=2e{\cal
  N}_0v_F\Delta_0$ as a unit for the current density. We should
mention that the current scales with the transparency of the contact,
${\cal T}_0$.  For other orientations of the order parameter,
$\alpha<22.5^\circ$, the same qualitative behavior should be observed
but with a reduced $T_\pi$.

Obviously, roughness suppresses the $\pi$-junction behavior (see
Fig.~\ref{TDCC}). With increasing $\tau$, the temperature $T_\pi$
decreases until the transition disappears for very rough interfaces.
This can be understood as follows: The negative current contributions
are carried by Andreev bound states with $E\lesssim 0$~\cite{Lo01}; as
they are broadened by disorder~\cite{Lu01} their negative current
contribution is partially canceled by those bound states with
$E\gtrsim 0$. Due to this suppression of the $\pi$-junction
contribution at rough interfaces, the normal $0$-junction contribution
becomes dominant also for lower temperatures; this contribution is
more stable against roughness as it is carried by bound states at
$E\gtrsim-|\Delta|$~\cite{Lo01}. For large enough disorder we finally
find $I_1>0$ in the whole temperature range. When further increasing
the roughness the critical current is enhanced until the negative
contributions vanish completely. Then, the temperature dependence of
the critical current, i.e. $I_c(T)/I_c(0.05T_c)$, has a monotonic
shape almost as in the $s$-wave case. For a very rough interface the
quantity $I_c(T)/I_c(0.05T_c)$ depends only weakly on the orientation,
$\alpha$ (see Fig.~\ref{UNC}).

Also with disorder the ground state at $T=T_\pi$ (as long as
$T_\pi>0$) exhibits a finite phase difference, which leads to a
spontaneous current parallel to the interface. As shown in
Fig.~\ref{PC} the current density increases rapidly with growing
interface roughness. This can be understood as follows: In the clean
case the directions preferring a $0$-junction and those preferring a
$\pi$-junction contribute to the current with opposite sign; this
leads to a partial cancellation also of the parallel current.  As
discussed before, roughness suppresses the $\pi$-junction behavior; as
a consequence the cancellation of the current contributions is less
effective, hence the current increases with roughness. It should be
mentioned that for this type of junction ($\alpha=22.5^\circ$) the
modification of the order parameter is small and plays only a minor
role.

The suppression of the $\pi$-junction behavior by disorder
has been reported earlier for alternative models of interface
roughness~\cite{Ba96,Po99}.

In the following, we will compare our results for the critical current
with the experimental data of Refs.~\cite{Di90,Hi98,Ma93,MaHi02}. There a
monotonic temperature dependence of the ratio $I(T)/I(0.05T_c)$ has
been reported, in agreement with our observations for very
rough interfaces.  The quantitative agreement of these results with
the data reported in~\cite{Hi98} for YBCO-junctions is reasonable for
small angles, $\alpha$, whereas for larger angles it is quite poor;
this can be seen when comparing the values of $R_NI_c$ at $T=4.2\;{\rm
K}\approx0.05\;T_c$ (see table~\ref{table}).
\begin{table}[b]
\begin{tabular}{p{1cm} p{2.1cm} p{1cm} p{1.8cm} p{1.8cm}}\\
$\alpha$ & $AR_N (\Omega {\rm cm}^2)$ & ${\cal T}_0^{\rm exp}$ &
$R_NI_c ({\rm mV})$ & $R_NI_c ({\rm mV})$\\
& \small{Ref.~\cite{Hi98}}
& & \small{Ref.~\cite{Hi98}} & \small{calculated}\\\hline
$12^\circ$ & $5.4\times 10^{-9}$ & $0.024$ & $1.3$ & $3.6$\\
$18^\circ$ & $1.5\times 10^{-8}$ & $0.008$ & $0.75$ & $2.6$\\
$22.5^\circ$ &$1.2\times 10^{-8}$ & $0.011$ & $0.13$ & $1.9$\\
\end{tabular}
\caption{\label{table}{Comparison of experimental results for $R_NI_c$
with calculated values. The transparency in the experiment, ${\cal
T}_0^{\rm exp}$, is estimated via Eq.~(\ref{res}) leading to the
values $AR_N$ given in the table, and $v_F=4.34\times 10^6{\rm cm/s}$,
${\cal N}_0=2.13\times 10^{22}/{\rm cm}^3{\rm eV}$~\cite{Ho93}. For the
calculations we used ${\cal T}_0=0.01$ (tunnel-limit), $T=0.05\;T_c$,
and $\tau=4$.}}
\end{table}
The discrepancy between our results and the experimental data might be
related to facets on $\mu{\rm m}$-scale at the
interface~\cite{Ma96,HiMa96}, which have an increasing influence for
larger angles $\alpha$, as more facets yield a negative current
contribution~\cite{HiMa96}.  An average over all facet orientations
then leads to a diminished total current.

Recently, for a small junction (width $0.5\mu{\rm m}$) with
$\alpha=22.5^\circ$ a non-monotonic temperature dependence of the
critical current has been reported~\cite{Il01} which, as already
discussed, appears due to the transition to a $\pi$-junction for low
temperatures. The minimum of the critical current was found for
$T_\pi=12\;{\rm K}=0.13\;T_c$. The transparency is quite high, as can
be seen from the large $I_2$-contribution. The measurement is in
qualitative agreement with results obtained from our model for ${\cal
T}_0\gtrsim 0.2$ and $\tau\gtrsim 0.4$. It is also worth mentioning
that this transition was observed for some of the samples only, which
might be due to the sensitivity of mesoscopic junctions to
fluctuations of the interface properties.  Note that here facets are
of minor importance, as the width of the junction is comparable to the
typical facet size; so it is possible to have junctions with
well-defined orientation.

In summary, we showed that the behavior of mirror junctions can be
modified drastically by interface roughness: If the roughness is weak
and the angle $\alpha$ is large enough, a transition to a
$\pi$-junction at low temperatures can be observed; near the
transition temperature, $T_\pi$, the junction has a
$\sin(2\varphi)$-like current-phase relation. For larger roughness the
$\pi$-junction behavior is destroyed and the quantity
$I_c(T)/I_c(0.05T_c)$ depends only weakly on the orientation
$\alpha$. When comparing with experimental data, we find qualitative
agreement with our theory.  However, especially the influence of large
facets has to be taken into account to determine the absolute value of
the critical current correctly.

\subsection{\label{AJ}$45^\circ$ asymmetric junction}
\begin{figure}[t]
\includegraphics[width=\linewidth]{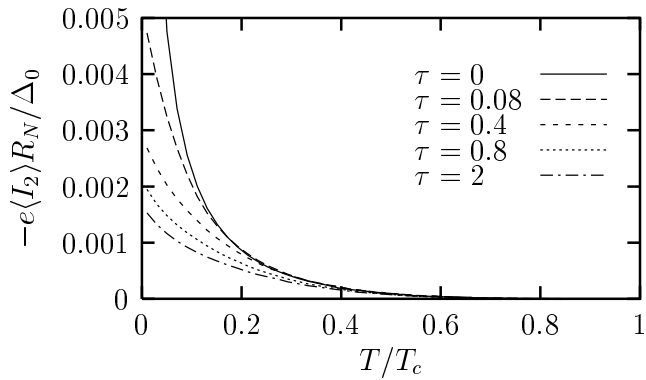}
\caption{\label{TDC2}Average current $I_2$ for an asymmetric junction
with ${\cal T}_0=0.01$ and varying strength of the roughness;
interface roughness suppresses the current $I_2$.}
\end{figure}
\begin{figure}[t]
\includegraphics[width=\linewidth]{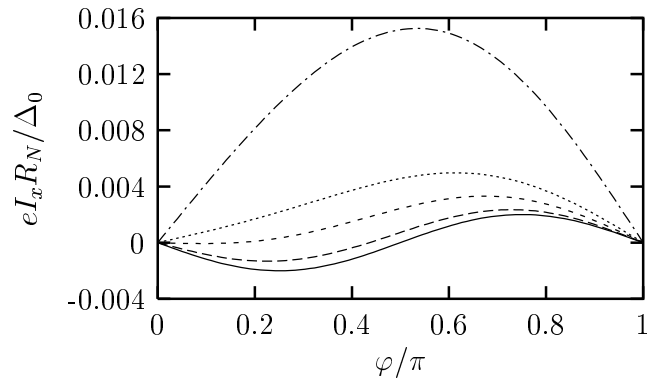}
\caption{\label{CPR1}Current-phase relation for single realizations of
the disorder for an asymmetric junction with $T=0.1T_c$. The
transparency is ${\cal T}_0=0.01$ and the roughness is given by the
same parameters $\tau$ as in Fig.~\ref{TDC2}.}
\end{figure}
\begin{figure}[t]
\includegraphics[width=\linewidth]{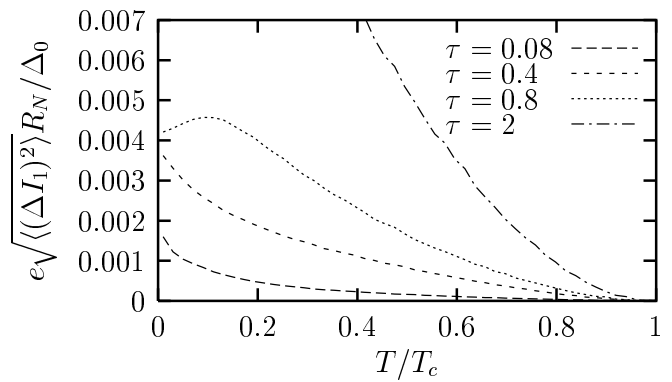}
\caption{\label{TDC1}Standard deviation of the current $I_1$ for an
asymmetric junction with ${\cal T}_0=0.01$ and varying strength of the
roughness, $\vartheta_c\approx\pi/40$; this current contribution is
increasing with the roughness.}
\end{figure}
\begin{figure}[t]
\includegraphics[width=\linewidth]{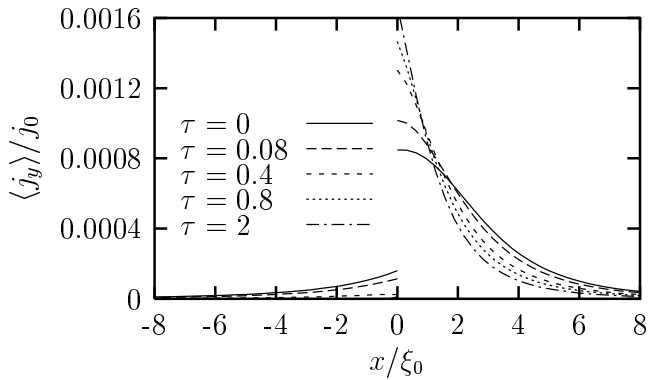}
\caption{\label{PCa}Average ground state (i.e. $\varphi\approx\pi/2$)
current density in $y$-direction at an asymmetric junction with ${\cal
T}_0=0.01$.}
\end{figure}

For a clean $45^\circ$ asymmetric junction, the particular geometric
symmetry is responsible for a vanishing leading current contribution
$I_1$: The symmetry operation $y\to -y$ leads to a phase-shift
$\varphi\to\varphi+\pi$, while on the other hand, the current in
$x$-direction must be invariant. We therefore find the following
condition for the current:
\begin{equation}\label{Iconf}
I_x(\varphi)=I_x(\varphi+\pi)\quad\Leftrightarrow\quad I_{2k-1}=0,
\quad k\in{\mathbb N},
\end{equation}
and the current-phase relation takes the form
\begin{equation} 
I_x(\varphi)=I_2\sin(2\varphi)+I_4\sin(4\varphi)+\dots.
\end{equation}
This means that in the absence of disorder the current is dominated by the
contribution $I_2\sin(2\varphi)$. Asymmetric junctions therefore have
two degenerate ground states at $\varphi=\pi/2,3\pi/2$. For this
reason such junctions are discussed as possible realizations of
quantum bits~\cite{Io99,Bl99}.

Considering rough interfaces we will first examine the average values
of $I_1$ and $I_2$; later we will also discuss the statistical
fluctuations and their relevance for physical realizations. We present
data for a tunnel-junction (${\cal T}_0=0.01$). Moreover in our
approach the statistical fluctuations depend on the angle
$\vartheta_c$ which is set to $\vartheta_c\approx\pi/40$ (i.e. $n=40$
directions were taken into account).

Since, in our model for rough interfaces, the symmetry $y\to -y$ is
still present on average, it follows that $\langle I_1\rangle=0$;
i.e. on average the tunnel-current exhibits a
$\sin(2\varphi)$-like current-phase relation.

The temperature dependence of $\langle I_2\rangle$ is shown in
Fig.~\ref{TDC2} for various roughness parameters. In the clean case
for ${\cal T}_0\to 0$ and $T\to 0$, we find $I_2\propto 1/T$ due to
the zero energy bound state at the interface; this state occurs as a
result of the sign change of the order parameter for reflected
quasi-particles on the right hand side.  But finite roughness also
leads to a suppression of this contribution and the average value
$\langle I_2\rangle$ decreases with growing interface roughness.  As
$\langle I_1\rangle=0$ the ground state of the junction has a
non-trivial phase difference $\varphi=\pi/2,3\pi/2$, and hence the
time-reversal symmetry is broken.  Thus a current parallel to the
interface exists (see Fig.~\ref{PCa}) whereas no current in
$x$-direction is present; as in the previous case the parallel current
scales with transparency.  The parallel current is carried by bound
states at $E\lesssim 0$, which exist on both sides of the interface
due to the finite transparency~\cite{Lo01}; the bound states with $E\gtrsim
0$ would contribute to the current in opposite direction, but they are
not occupied.  Interface roughness generally leads to a suppression of
the parallel current since the bound states are broadened, and their
current contributions tend to cancel each other. Only in a narrow
region on the right hand side of the interface, of the order of the
coherence length, the parallel current is enhanced, which is related
to the considerable increase of the tilted order parameter at a rough
interface~\cite{Lu01}.

The average critical current of $45^\circ$ asymmetric junctions was
considered before for the clean~\cite{TaKa97,Hu97,Fo98} as well as for
the rough case using the thin dirty layer model~\cite{Bu97}. The
results are similar to ours. The spontaneous parallel current was
studied for the clean case~\cite{Fo98}.  Roughness was also considered
for a completely transparent interface~\cite{Am02}, where a reduction
of the parallel current was found for increasing roughness. The
$\sin(2\varphi)$-like behavior has been observed in an experiment with
a YBCO-junction by Il'ichev et al.~\cite{Il99}.

So far we assumed that the junction is well described by the average
values of the current, but when considering a particular realization
of the disorder (i.e. of the scattering matrix) the symmetry stated in
Eq.~(\ref{Iconf}) is no longer valid, and a finite contribution $I_1$
is expected. To confirm this assertion we consider the current-phase
relation for one particular scattering matrix at the temperature
$T=0.1\;T_c$; the result is shown in Fig.~\ref{CPR1}. The
$\sin(2\varphi)$-behavior, observable in the clean case, is clearly
modified by the roughness and a strong $\sin\varphi$-part additionally
occurs; for $\tau\gtrsim 0.4$ the current-phase relation is already
dominated by this contribution.  In order to study the statistical
properties of the current more systematically we consider the standard
deviation of the contribution $I_1$, namely $\langle(\Delta
I_1)^2\rangle^{1/2}$; its temperature dependence is shown in
Fig.~\ref{TDC1}. Even for small roughness the quantity $\langle(\Delta
I_1)^2\rangle^{1/2}$ may already be of the same order of magnitude as
$\langle I_2\rangle$, in particular for higher temperatures (compare
Figs.~\ref{TDC2} and~\ref{TDC1}). The reason is the different
dependence on the transparency of both contributions, $I_1\propto{\cal
T}_0$ and $I_2\propto{\cal T}_0^2$, so that for ${\cal T}_0\ll 1$ the
contribution $I_1$ dominates as soon as it is allowed by symmetry.

Recalling the physical meaning of the averaging process (see
Sec.~\ref{SM}) we conclude that the statistical fluctuations should be
relevant for small junctions ($A$ of the order of $\xi_0^2$ or even
smaller), and a $\sin\varphi$-like current phase relation should be
observable. On the other hand for large junctions ($A\gg\xi_0^2$) the
statistical fluctuations become irrelevant, and the behavior should be
governed by the $\sin(2\varphi)$-like contribution. As already
mentioned above a $\sin(2\varphi)$-like current-phase relation has
been observed in experiment~\cite{Il99}. But only some of the samples
showed this behavior; statistical fluctuations as discussed here can
be ruled out as a reason, however, because the contact in the
experiment is quite large compared to the coherence length of YBCO,
$\xi_0\approx 15\;{\rm\AA}$: $A\approx 1\;\mu{\rm m}\times 100\;{\rm
nm}$.  A reasonable explanation might be a slight deviation from the
$45^\circ$-orientation of the tilted order parameter in some samples,
such that the contribution $I_1$ is allowed by symmetry. Another
explanation might be a faceted interface, so that for each facet the
contribution $I_1$ is finite.

\section{\label{CO}Conclusion}

To describe a contact between two superconductors within the
quasiclassical theory, we adapted the boundary conditions suggested
in~\cite{ShOz00}.  For numerical calculations the choice of a discrete
grid of directions $\Fe p$ is of particular importance; we have shown
how this grid must be chosen in order to guarantee current
conservation perpendicular to the interface for an arbitrary
scattering matrix. Thereafter we proposed a phenomenological random
scattering matrix in order to describe microscopically rough interfaces
without regular structure.

In section~\ref{RD} we applied our model to mirror junctions and
$45^\circ$ asymmetric junctions with interface roughness. At first we
studied the evolution of the temperature-dependent critical current
for mirror junctions when increasing the interface roughness: We found
that the non-monotonic behavior for a specular interface turns to a
monotonic temperature dependence for rough interfaces. Therefore
interface roughness is a possible explanation for the experimental
results.

For the asymmetric junction we calculated the average current-phase
relation, which describes large contacts ($A\gg\xi_0^2$). We find a
leading $\sin(2\varphi)$-like behavior of the current phase relation.
This is in agreement with experimental findings~\cite{Il99} as
discussed in Sec.~\ref{AJ}. On the other hand we also showed that in
small enough junctions a dominating $\sin\varphi$-like contribution
might be present. It would be interesting to check this effect
experimentally.

In conclusion our results explain several aspects of the experimental
data. In order to improve the theoretical description, a detailed
description of the interface is necessary in the first place. For
example facets, or a modified electronic structure near the interface,
should be taken into account in a more realistic theory~\cite{MaHi02}.
\section*{Acknowledgments}
We would like to thank J. Mannhart and M. Ozana for stimulating
discussions. This work was supported in part by the German Research
Foundation (DFG).

\newpage
\bibliography{manuall}

\end{document}